\documentclass[%
 reprint,
 amsmath,amssymb,
 aps,
pra,
]{revtex4-1}

\usepackage{float}
\usepackage[usenames, dvipsnames]{color}
\usepackage{graphicx}
\usepackage{dcolumn}
\usepackage{bm}

\begin{document}

\preprint{APS/123-QED}

\title{Effects of Field Fluctuations on Driven Autoionizing
Resonances}

\author{G. Mouloudakis$^1$}
 \email{gmouloudakis@physics.uoc.gr}

\author{P. Lambropoulos$^{1,2}$}%

\affiliation{${^1}$Department of Physics, University of Crete, P.O. Box 2208, GR-71003 Heraklion, Crete, Greece
\\
${^2}$Institute of Electronic Structure and Laser, FORTH, P.O.Box 1527, GR-71110 Heraklion, Greece}

\date{\today}

\begin{abstract}

The excitation of an autoionizing resonance by intense radiation requires a theoretical description beyond the transition probability per unit time. This implies a time-dependent formulation incorporating all features of the radiation source, such as pulse temporal shape and duration, as well as stochastic properties, for pulses other than Fourier limited. The radiation from short wavelength free electron lasers is a case in point, as it is the only source that can provide the necessary intensity. In view of ongoing experiments with such sources, we present a systematic study for an isolated autoionizing resonance. We find that intensity, pulse duration and field fluctuations conspire in producing unexpected excitation profiles, not amenable to a description in terms of the usual Fano profile. In particular, the role of intensity fluctuations turns out to pose challenging theoretical problems part of which have been addressed herein.  

\begin{description}

\item[PACS numbers]
32.80.Aa, 32.80.Hd, 32.70.Jz, 32.80.Rm

\end{description}
\end{abstract}

\pacs{Valid PACS appear here}
\maketitle


\section{\label{sec:level1}Introduction}

Autoionization is a process by which an atom in an excited state spontaneously ejects one of its electrons due to its interaction with a continuum \cite{ref1}. A common example of an autoionizing state (AIS) - also referred to as AI resonance - is a discrete state involving the excitation of two electrons with total energy larger than the one-electron ionization threshold. Such states are unstable and will eventually decay non-radiatevely due to their interaction with the continuum. If the width of the excitation profile of an AIS, is much smaller than the energy distance from the nearest AIS, it is usually referred to as  an isolated AIS. The paradigm of an isolated AIS is provided by the doubly excited 2s2p state of Helium which can be excited by radiation in the XUV range; with photon energy around 60 eV.

In most traditional photoabsorption studies of AIS', the resonances were excited by synchrotron sources with low intensities and practically monochromatic radiation. The availability of strong radiation sources, such as lasers, motivated the exploration of the behavior of resonant transitions, driven by strong, pulsed and possibly non-monochromatic radiation. Over the last 35 years or so, a plethora of related studies have addressed issues such as AC Stark splitting in strongly driven bound states in double optical resonance, including the effect of field fluctuations \cite{ref2,ref3}. Those studies were limited to the optical or near UV spectral region, in which sources of sufficient intensity were at the time available. The strong driving of AI states, such as the 2s2p in Helium, which requires radiation in the XUV range, were beyond the reach of those sources. Nevertheless, some initial theoretical exploration of the expected behavior of a strongly driven AI state, as well as the case of double resonance, involving the strong coupling of  two AI states were published as early as 1980 \cite{ref4,ref5} and revisited much later \cite{ref6}, when short wavelength Free Electron Lasers (FEL) began delivering strong radiation in the XUV and beyond. Extension of that work to triply excited hollow states followed a few years later \cite{ref7,ref8,ref9}.  To the best of our knowledge, there are two examples \cite{ref10,ref11} of experimental work providing some evidence of the strong coupling of two AI states. 

At this point, we need to define the notion of strong coupling in AI states. A formulation of an AI state, particularly valid for an isolated resonance, involves the superposition of a discrete state and the continuum to which it is coupled via intra-atomic interaction. Diagonalization of the relevant part of the Hamiltonian leads to a modification of the position of the discrete part and a decay rate, referred to as AI width, as it corresponds to the width of the profile of the resonance. As discussed in detail in the next section, the dipole matrix element coupling a bound state to the discrete part of the AI resonance multiplied by the electric field amplitude, to within some coefficients, represents an effective Rabi frequency characterizing the strength of the coupling. Under traditional synchrotron radiation experiments, that Rabi frequency in much smaller than the AI width. That is what we shall call weak coupling, in which case the interaction is describable in terms of a transition probability per unit time (rate), as given by Fermi's golden rule. The coupling is strong when the above condition is reversed, with the Rabi frequency being larger than the AI width. When an EM field couples two AI states, as for example in refs. \cite{ref4,ref5,ref6}, the field is strong when the Rabi frequency between the two AI states is larger than the AI width of at least one of the resonances.

The availability of strong, short pulse duration, short wavelength (XUV and beyond) radiation through the recent Free Electron Laser (FEL) sources \cite{ref12} provides the opportunity to explore the behavior of strongly driven AI resonances. For a pulse of high peak intensity, the pulse duration enters as an important parameter paired to the intensity. Clearly, under any intensity, given sufficiently long time exposure of any system to the radiation, complete ionization will ensue. When the intensity is high, even a seemingly short pulse duration can cause significant ionization, to the extent of distorting the resonance profile. That effect noted in the early paper by Lambropoulos and Zoller \cite{ref5} and referred to as "time saturation", has until now been of only academic interest. But as shown in this paper, it should be expected to be of crucial influence in experiments under FEL radiation.
 
In addition to the high intensity and short pulse duration, at least at the present time, FEL’s exhibit strong intensity fluctuations \cite{ref12}. This means that the radiation seen by the atom has a non-zero bandwidth, while the Rabi frequency will undergo stochastic fluctuations. The influence of field fluctuations has not received much attention in the theoretical literature, until now, simply because AI resonances tend to be much broader than the bandwidths of synchrotron sources. Under FEL radiation, however, that is no longer the case, which requires a formulation that accounts for the stochastic fluctuations of the field. Depending on the strength of the coupling to the radiation, it may be that only the bandwidth is of importance. In the most general case of strong driving, in the sense defined above, accounting for the intensity fluctuations becomes imperative.

As demonstrated in the sections that follow, hitherto "academic" effects such as time broadening are expected to cause dramatic distortion of the profile. Moreover the complete theory of an AI resonance driven strongly by a stochastic field, with intensity fluctuations, poses challenging problems not readily amenable to the theoretical tools developed for strongly driven transitions between bound states \cite{ref2}. In a recent short paper \cite{ref13}, we have reported a first assessment of the basic effects to be expected. The present paper provides a more detailed exposition of the theory, as well as a partial account of the effect of intensity fluctuations.  In the light of our treatment and results,  a number of important aspects, outlined in the concluding remarks, pose challenging open problems.

A few comments on the features of the FEL radiation are in order, at this point. The source is pulsed. The term pulse duration means the full width at half maximum of the temporal profile, while the peak intensity will often be referred to simply as the intensity. If in comparing theory to experimental data, we were to be concerned with accuracy of 1$\% $ or better, then the exact temporal profile would be of importance. But given the present state of uncertainty in the source parameters, long experience has shown that assuming a Gaussian temporal profile is more than adequate. Moreover, for illustrative purposes of various aspects of the theory, again experience has shown that assuming a square pulse of roughly the same duration does encapsulate the essential physics. Its occasional use in this paper is limited to the cases in which analytic feasibility of the theory is deemed informative.

\section{\label{sec:level2}Theoretical Formulation}
We begin by considering a two-level atom whose ground state $\left| 1 \right\rangle$ is coupled to an isolated autoionizing state (AIS) $\left| 2 \right\rangle$ via a single-photon transition, in the presence of an external electric field $E(t) = \frac{1}{2}[\mathcal{E} (t){e^{i\omega t}} + c.c.]$. Its frequency $\omega$ is tuned around the resonant frequency ${\omega _{21}} \equiv {\omega _2} - {\omega _1}$ and $\mathcal{E} (t) = \left| {\mathcal{E} (t)} \right|\exp [i\varphi (t)]$ is in general assumed to undergo stochastic fluctuations. The coupling to the resonance is characterized by the complex Rabi frequency $\tilde \Omega (t) = \Omega (t)(1 - \frac{i}{q}) = \frac{1}{2}\mathcal{E} (t){d_{21}}(1 - \frac{i}{q})$, where ${d_{21}}$ is the electric dipole matrix element between the ground state and the discrete part of the resonance and q the asymmetry parameter \cite{ref14}. The ground state is also coupled directly to the continuum via a dipole matrix element accounting for ionization into the smooth continuum, leading to an ionization width denoted by $\gamma (t)$. The autoionization width $\Gamma$ represents the rate of decay of the AI state due to the Coulomb interaction between the two excited electrons. Spectroscopically, it appears as the width of the excitation profile of the resonance and is equal to the inverse of the lifetime of the AIS. The q parameter accounts for the interference between the two paths to the continuum; the direct and the one via the discrete part, and is related to the Rabi frequency and the two widths through the strict relation $4{\Omega ^2} = {q^2}\gamma \Gamma $.

Depending on the aspects of the problem to be  addressed, the theory can be cast either in terms of the time-dependent Schrondinger equation or the density matrix. In what follows, we treat the problem in terms of the density matrix $\rho(t)$ because it allows the distinction between phase and intensity fluctuations, which is instructive. It provides in addition a convenient  tool for useful approximations, such as the decorrelation of atomic and field dynamics.

The dynamical evolution of the  slowly varying part ${\sigma}(t)$ of the density matrix is governed by the following equations \cite{ref15,ref16}:  
\begin{equation}
{\partial _t}{\sigma _{11}}(t) =  - \gamma (t){\sigma _{11}}(t) + 2Im\left\{ {\Omega (t)\left( {1 - \frac{i}{q}} \right){\sigma _{21}}(t)} \right\}
\end{equation}
\begin{equation}
{\partial _t}{\sigma _{22}}(t) =  - \Gamma {\sigma _{22}}(t) - 2Im\left\{ {\Omega (t)\left( {1 + \frac{i}{q}} \right){\sigma _{21}}(t)} \right\}
\end{equation}
\begin{equation}
\left[ {{\partial _t} - i\Delta  + \frac{1}{2}\left( {\gamma (t) + \Gamma } \right)} \right]{\sigma _{21}}(t) =  - i\Omega (t)\left( {1 - \frac{i}{q}} \right){\sigma _{11}}(t) $$
$$ + i\Omega (t)\left( {1 + \frac{i}{q}} \right){\sigma _{22}}(t)
\end{equation}
\\
where we have introduced the slowly varying matrix elements ${\sigma _{ij}}(t)$, defined by ${\rho _{ii}}(t) = {\sigma _{ii}}(t)$ , $i = 1,2$ and ${\rho _{21}}(t) = {\sigma _{21}}(t)exp[i\omega t]$. The detuning $\Delta$ of the photon frequency from resonance is defined by $\Delta  \equiv \omega  - {\omega _{21}}$. Note that the left side of equation (3) may in general contain additional coherence (off-diagonal) relaxation constants which are of no relevance to our problem, in the case of monochromatic field. However, a coherence relaxation constant appears below, as we introduce field fluctuations.

The matrix elements of the density matrix in the above equations are generally fluctuating variables owing to the stochastic character of the field which imparts fluctuations to the  Rabi frequency and the ionization width. We are therefore dealing with stochastic differential equations. The observed quantities which refer to the atom are given by the average over the stochastic fluctuations of the field. This requires a realistic model of the stochastic properties of the field, or a brute force numerical integration over trajectories imitating the fluctuations of the field.
We have here adopted the first approach. 

To this end, we solve equation (3) for ${\sigma _{21}}(t)$ formally, and substitute into equations (1) and (2). Denoting the stochastic averages of the resulting equations by angular brackets, we obtain:
\begin{widetext}

\begin{equation}
{\partial _t}\left\langle \sigma _{11}(t) \right\rangle = 
- \left\langle \gamma (t) \sigma _{11} (t) \right\rangle  
+ 2{\rm Im}  \left\{ \left (1 - \frac{i}{q} \right )
\int_0^t -i \left (1 - \frac{i}{q} \right )  
\left\langle {\Omega (t)\Omega (t') \sigma _{11}(t')} \right\rangle 
e^{ - \kappa (t - t')} dt' \right. $$
$$ +\left. \left (1 - \frac{i}{q} \right ) \int_0^t i\left (1 + \frac{i}{q} \right )
\left\langle \Omega (t)\Omega (t')\sigma _{22} (t') \right\rangle 
e^{ - \kappa (t - t')}dt'  \right\} 
\end{equation}

\begin{equation}
{\partial _t}\left\langle \sigma _{22} (t) \right\rangle =  
 - \Gamma \left\langle \sigma _{22}(t) \right\rangle  
 - 2{\rm Im} \left\{ \left (1 + \frac{i}{q}\right )
\int_0^t -i \left (1 - \frac{i}{q} \right )  
\left\langle {\Omega (t)\Omega (t') \sigma _{11}(t')} \right\rangle 
e^{ - \kappa (t - t')} dt' \right. $$
$$ +\left. \left (1 + \frac{i}{q} \right ) \int_0^t i\left (1 + \frac{i}{q} \right )
\left\langle \Omega (t)\Omega (t')\sigma _{22} (t') \right\rangle 
e^{ - \kappa (t - t')}dt'  \right\} 
\end{equation}

where to compress notation somewhat,we have introduced   $\kappa  \equiv  - i\Delta  + \frac{1}{2}(\gamma  + \Gamma )$.

\end{widetext}

Equations (4) and (5) involve atom-field correlation functions of the form $\left\langle {\Omega (t)\Omega (t'){\sigma _{ii}}(t')} \right\rangle $, $i = 1,2$. Generally such correlation functions cannot be evaluated without knowing the specific form of the fluctuations of the field. As an approximation valid under certain conditions, one could decorrelate the atomic-field dynamics \cite{ref2} by taking $\left\langle {\Omega (t)\Omega (t'){\sigma _{ii}}(t')} \right\rangle  = \left\langle {\Omega (t)\Omega (t')} \right\rangle \left\langle {{\sigma _{ii}}(t')} \right\rangle $. There are, however, specific models of fluctuating fields where the decorrelation is mathematically rigorous. In the following, we describe briefly two widely used models and apply them to the context of our problem; namely, the phase-diffusion and the chaotic field model.
\\

In the phase-diffusion (PD) model, the field has a non-fluctuating amplitude but its phase is a Wiener-Levy stochastic process [R]. In that case the nth-order correlation function of the field obeys the relation \cite{ref18}
\begin{equation}
\left\langle {{\varepsilon ^*}({t_1})\varepsilon ({t_2})...{\varepsilon ^*}({t_{2n - 1}})\varepsilon ({t_{2n}})} \right\rangle  = \prod\limits_{j \to odd}^{2n - 1} {\left\langle {{\varepsilon ^*}({t_j})\varepsilon ({t_{j + 1}})} \right\rangle }
\end{equation}
with ${t_j} > {t_{j + 1}}$.
This represents a Markovian process, with an exponential first-order correlation function given by \cite{ref17}
\begin{equation}
\left\langle {{\varepsilon ^*}({t_1})\varepsilon ({t_2})} \right\rangle  = \left\langle {{{\left| {\varepsilon (t)} \right|}^2}} \right\rangle \exp [ - \frac{1}{2}{\gamma _L}\left| {{t_1} - {t_2}} \right|]
\end{equation}
where ${\gamma _L}$ is the bandwidth of the field.

It has been established \cite{ref2} that, in the case of the phase-diffusion model, the decorrelation of the atom-field dynamics is rigorous without any approximation. Physically, this is easy to understand, because for a constant amplitude, the fluctuations of the phase of the field cannot affect the evolution of the populations, but only the coherence, which means the relative phase of the coefficients representing the superposition of the states coupled by the field. And it is the correlation between the time evolution of populations that is factorized out in the process of decorrelation. As we will discuss shortly, this is not the case when also the amplitude undergoes random fluctuations.

Formally, in our problem, the decorrelation implies the  relation  $\left\langle {\Omega (t)\Omega (t'){\sigma _{ii}}(t')} \right\rangle  = \left\langle {\Omega (t)\Omega (t')} \right\rangle \left\langle {{\sigma _{ii}}(t')} \right\rangle$. Note that ${\Omega (t)}$ is real. The same argument as above justifies the decorrelation of the quantities in the term $\left\langle {\gamma (t){\sigma _{11}}(t)} \right\rangle $, because
${\gamma (t)}$, representing the direct single-photon ionization from the ground state, is proportional to the intensity which does not undergo fluctuations . 
\\

In the chaotic field model the field undergoes both amplitude and phase fluctuations. Its amplitude is a complex Gaussian stochastic process with its nth order correlation function obeying \cite{ref18}:
\begin{equation}
\left\langle {{\varepsilon ^*}({t_1})\varepsilon ({t_2})...{\varepsilon ^*}({t_{2n - 1}})\varepsilon ({t_{2n}})} \right\rangle  = \sum\limits_P {\prod\limits_{j \to odd}^{2n - 1} {\left\langle {{\varepsilon ^*}({t_j})\varepsilon ({t_{P(j + 1)}})} \right\rangle } } 
\end{equation}
where the sum is over all possible permutations P, with ${t_j} > {t_{j + 1}}$.
For the sake of simplicity we will assume the chaotic field to be Markovian, something that it is not necessarily satisfied for a general chaotic field. In that case the first-order correlation function of the field is given by equation (7).

In contrast to the case of the phase diffusion model, the decorrelation
$\left\langle {\Omega (t)\Omega (t'){\sigma _{ii}}(t')} \right\rangle  = \left\langle {\Omega (t)\Omega (t')} \right\rangle \left\langle {{\sigma _{ii}}(t')} \right\rangle$ is not mathematically rigorous for a chaotic field, but as an approximation, it is valid in the weak field regime. The relative errors of this approximation have been evaluated recently, as a function of the ratio $\Omega /\Gamma $ for various field bandwidths \cite{ref19}. It has been shown that the error of the decorrelation approximation (DA) becomes significant for increasing intensities and laser bandwidths comparable to the autoionization width. The variables $\gamma (t)$ and ${\sigma _{11}}(t)$ are also decorrelated within the DA since $\gamma (t)$ is proportional to the intensity which is approximately replaced by its averaged value.

In view of the above, we proceed with the decorrelation of the atomic-field dynamics, with the resulting equations being exact for the PD model and valid in the weak to moderate field limit for the chaotic model. After the decorrelation, equations (4) and (5) become:
\begin{widetext}

\begin{equation}
{\partial _t}\left\langle {{\sigma _{11}}(t)} \right\rangle  =  -\left\langle \gamma (t)\right\rangle\left\langle {{\sigma _{11}}(t)} \right\rangle  + d_{12}^2I(t){\mathop{\rm Im}\nolimits} \left\{ \left( {1 - \frac{i}{q}} \right)\int\limits_0^t  - i\left( {1 - \frac{i}{q}} \right)\left\langle {{\sigma _{11}}(t')} \right\rangle {e^{ - \tilde \kappa (t - t')}}dt' \right. $$
$$+\left. \left( {1 - \frac{i}{q}} \right)\int\limits_0^t {i\left( {1 + \frac{i}{q}} \right)\left\langle {{\sigma _{22}}(t')} \right\rangle {e^{ - \tilde \kappa (t - t')}}dt'}  \right\}
\end{equation}

\begin{equation}
{\partial _t}\left\langle {{\sigma _{22}}(t)} \right\rangle  =  - \Gamma \left\langle {{\sigma _{22}}(t)} \right\rangle  - d_{12}^2I(t){\mathop{\rm Im}\nolimits} \left\{ \left( {1 + \frac{i}{q}} \right)\int\limits_0^t  - i\left( {1 - \frac{i}{q}} \right)\left\langle {{\sigma _{11}}(t')} \right\rangle {e^{ - \tilde \kappa (t - t')}}dt' \right. $$
$$+\left. \left( {1 + \frac{i}{q}} \right)\int\limits_0^t {i\left( {1 + \frac{i}{q}} \right)\left\langle {{\sigma _{22}}(t')} \right\rangle {e^{ - \tilde \kappa (t - t')}}dt'}  \right\}
\end{equation}

\end{widetext}

where we have substituted the complete expression of the Rabi frequency $ \Omega (t)= \frac{1}{2}\mathcal{E} (t){d_{21}}$ and defined $\tilde \kappa  \equiv \kappa  + \frac{1}{2}{\gamma _L} =  - i\Delta  + \frac{1}{2}(\gamma  + \Gamma  + {\gamma _L})$. In the RWA, the intensity appearing in equations (9) and (10) is expressed in terms of the field amplitude as $I(t) = \frac{{\left\langle {{{\left| {\mathcal{E}(t)} \right|}^2}} \right\rangle }}{2}$.

Let us consider, for the moment, the case of constant intensity $I(t) = {I_0}$, which leads to considerable simplification enabling  analytical solutions. Since the integrals appearing in (8) and (9) are with respect to the time t, which is a real variable, the interchange of the Imaginary (or Real) part and integration are mathematically rigorous. Using this fact and expanding the exponential functions in the integrands in terms of Cosine and Sine functions, we obtain:
\begin{widetext}

\begin{equation}
{\partial _t}\left\langle {{\sigma _{11}}(t)} \right\rangle  =  - \gamma \left\langle {{\sigma _{11}}(t)} \right\rangle  + d_{12}^2{I_0}\left( { - 1 + \frac{1}{{{q^2}}}} \right)\int\limits_0^t {\left\langle {{\sigma _{11}}(t')} \right\rangle {e^{ - \frac{1}{2}(\gamma  + \Gamma  + {\gamma _L})(t - t')}}} \cos [\Delta (t - t')]dt' $$
$$ + d_{12}^2{I_0}\left( { - \frac{2}{q}} \right)\int\limits_0^t {\left\langle {{\sigma _{11}}(t')} \right\rangle {e^{ - \frac{1}{2}(\gamma  + \Gamma  + {\gamma _L})(t - t')}}} \sin [\Delta (t - t')]dt' + d_{12}^2{I_0}\left( {1 + \frac{1}{{{q^2}}}} \right)\int\limits_0^t {\left\langle {{\sigma _{22}}(t')} \right\rangle {e^{ - \frac{1}{2}(\gamma  + \Gamma  + {\gamma _L})(t - t')}}} \cos [\Delta (t - t')]dt'
\end{equation}
\begin{equation}
{\partial _t}\left\langle {{\sigma _{22}}(t)} \right\rangle  =  - \Gamma \left\langle {{\sigma _{22}}(t)} \right\rangle  - d_{12}^2{I_0}\left( { - 1 - \frac{1}{{{q^2}}}} \right)\int\limits_0^t {\left\langle {{\sigma _{11}}(t')} \right\rangle {e^{ - \frac{1}{2}(\gamma  + \Gamma  + {\gamma _L})(t - t')}}} \cos [\Delta (t - t')]dt' $$
$$- d_{12}^2{I_0}\left( { - \frac{2}{q}} \right)\int\limits_0^t {\left\langle {{\sigma _{22}}(t')} \right\rangle {e^{ - \frac{1}{2}(\gamma  + \Gamma  + {\gamma _L})(t - t')}}} \sin [\Delta (t - t')]dt'- d_{12}^2{I_0}\left( {1 - \frac{1}{{{q^2}}}} \right)\int\limits_0^t {\left\langle {{\sigma _{22}}(t')} \right\rangle {e^{ - \frac{1}{2}(\gamma  + \Gamma  + {\gamma _L})(t - t')}}} \cos [\Delta (t - t')]dt'
\end{equation}

\end{widetext}

Note that for constant intensity, the width of the direct ionization to the continuum is also constant, i.e $\gamma (t) = \gamma $.
The integrals appearing in equations (10) and (11) are now convolutions of $\left\langle {{\sigma _{ii}}(t')} \right\rangle $ , $i = 1,2$ and the Sin/Cosine functions. Taking now the Laplace transforms of the above equations we obtain:

\begin{equation}
s{F_1}(s) =  - \gamma {F_1}(s) + d_{12}^2{I_0}\left[ \left(  - 1 + \frac{1}{{{q^2}}} \right){F_1}(s){G_1}(s) \right. $$
$$ \left. - \frac{2}{q}{F_1}(s){G_2}(s)+\left( 1 + \frac{1}{{{q^2}}} \right){F_2}(s){G_1}(s) \right]
\end{equation}
\begin{equation}
s{F_2}(s) =  - \Gamma {F_2}(s) - d_{12}^2{I_0}\left[ \left(  - 1 - \frac{1}{{{q^2}}} \right){F_1}(s){G_1}(s) \right. $$
$$ \left. - \frac{2}{q}{F_2}(s){G_2}(s)+\left( 1 - \frac{1}{{{q^2}}} \right){F_2}(s){G_1}(s) \right]
\end{equation}

where with ${{F_1}(s)}$ and ${{F_2}(s)}$ we denote the Laplace transforms of ${\left\langle {{\sigma _{11}}(t)} \right\rangle }$ and ${\left\langle {{\sigma _{22}}(t)} \right\rangle }$, while with ${{G_1}(s)}$ and ${{G_2}(s)}$ the Laplace transforms of the functions ${{g_1}(t) = {e^{ - \frac{1}{2}(\gamma  + \Gamma  + {\gamma _L})t}}\cos (\Delta t)}$ and ${{g_2}(t) = {e^{ - \frac{1}{2}(\gamma  + \Gamma  + {\gamma _L})t}}sin(\Delta t)}$, respectively. It is a matter of straightforward algebraic manipulations to show that
\begin{equation}
{{G_1}(s) = \frac{{s + \frac{1}{2}(\gamma  + \Gamma  + {\gamma _L})}}{{{{\left[ {s + \frac{1}{2}(\gamma  + \Gamma  + {\gamma _L})} \right]}^2} + {\Delta ^2}}}}
\end{equation}
and
\begin{equation}
{{G_2}(s) = \frac{\Delta }{{{{\left[ {s + \frac{1}{2}(\gamma  + \Gamma  + {\gamma _L})} \right]}^2} + {\Delta ^2}}}}
\end{equation}
\\
The system of equations (12) and (13) can be solved easily for ${{F_1}(s)}$ and ${{F_2}(s)}$, the inverse Laplace transform of which provide the exact time dependence of $\left\langle {{\sigma _{11}}(t)} \right\rangle $ and $\left\langle {{\sigma _{22}}(t)} \right\rangle $. The expressions are too lengthy and complicated to be visually enlightening,  but the results are discussed in later sections.

Returning now to the more general case of time-dependent intensity $I(t)$, the above analytical treatment using the Laplace transform does not lead to helpful expressions and even the numerical solution of equations (8) and (9) tends to be a very cumbersome task. However, useful insight can be gained through the approximation, 
\begin{equation}
\begin{array}{*{20}{c}}
{\left\langle {{\sigma _{ii}}(t')} \right\rangle  \simeq \left\langle {{\sigma _{ii}}(t' = t)} \right\rangle }&{}&{i = 1,2}
\end{array}
\end{equation}
which is valid in the weak field limit. Its validity rests upon the realization that, under weak driving, the populations do not change significantly, over times of rapid oscillations of the rest of the integrand. As a result they can be evaluated at times $t' = t$ and factored out of the integral. In that case, the integration with respect to $t'$ in equations (8) and (9) can be performed, leading to a somewhat simplified system of differential equations governing the time evolution of the populations, known as rate equations. They are:

\begin{widetext}

\begin{equation}
{\partial _t}\left\langle {{\sigma _{11}}(t)} \right\rangle  =  - \gamma (t)\left\langle {{\sigma _{11}}(t)} \right\rangle  + d_{12}^2I(t){\mathop{\rm Im}\nolimits} \left\{ {\frac{{1 - {e^{ - \tilde \kappa t}}}}{{\tilde \kappa }}\left[ {\left( { - i} \right){{\left( {1 - \frac{i}{q}} \right)}^2}\left\langle {{\sigma _{11}}(t)} \right\rangle  + i\left( {1 - \frac{i}{q}} \right)\left( {1 + \frac{i}{q}} \right)\left\langle {{\sigma _{22}}(t)} \right\rangle } \right]} \right\}
\end{equation}
\begin{equation}
{\partial _t}\left\langle {{\sigma _{22}}(t)} \right\rangle  =  - \Gamma \left\langle {{\sigma _{22}}(t)} \right\rangle  - d_{12}^2I(t){\mathop{\rm Im}\nolimits} \left\{ {\frac{{1 - {e^{ - \tilde \kappa t}}}}{{\tilde \kappa }}\left[ {\left( { - i} \right)\left( {1 - \frac{i}{q}} \right)\left( {1 + \frac{i}{q}} \right)\left\langle {{\sigma _{11}}(t)} \right\rangle  + i{{\left( {1 + \frac{i}{q}} \right)}^2}\left\langle {{\sigma _{22}}(t)} \right\rangle } \right]} \right\}
\end{equation}

\end{widetext}

The system of equations (17) and (18) can now be solved numerically in the limit of weak to moderate fields, as long as the Rabi frequency is less than or comparable to the autoionization width $\Gamma$. Actually, for special forms of the time dependent $I(t)$, analytical solutions may also be obtained.

\section{\label{sec:level3}Results and Discussion}
In this section we present and discuss the main results of this work on an AIS driven by stochastically fluctuating fields. For a quantitative analysis, we apply our theory to the case of Helium 2s2p ${}^1P$  AIS which offers a perfect example of an isolated autoionizing resonance. 

We begin by solving the system of equations (13) and (14) and inverting the Laplace transforms to obtain the expressions for $\left\langle {{\sigma _{11}}(t)} \right\rangle $ and $\left\langle {{\sigma _{22}}(t)} \right\rangle $. The ionization probability at the end of the square pulse (constant intensity) at a time T would normally be given by ${P_{ion}}(T) = 1 - \left\langle {{\sigma _{11}}(T)} \right\rangle  - \left\langle {{\sigma _{22}}(T)} \right\rangle $. However, at time T there will be population in the excited state that will decay to the continuum with a rate $\Gamma $ due to the configuration interaction (the spontaneous decay rate is negligible compared to $\Gamma $). Therefore, in order to account for this population we should express the ionization probability at times $t > T$ as \cite{ref5}:
\begin{equation}
{P_{ion}}(t) = 1 - \left\langle {{\sigma _{11}}(t)} \right\rangle  - \left\langle {{\sigma _{22}}(t)} \right\rangle {e^{ - \Gamma (t - T)}}
\end{equation}

We can now plot the ionization probability calculated at times $t > T$ as a function of the driving frequency around the resonance, for various intensities, laser bandwidths and  interaction times T.
However, for this particular AIS, in addition to the ionization of the neutral, the radiation can ionize the $ {He(1s)}^{+}$ ions produced from autoionization. This  process involves the absorption  of one additional photon, producing $He^{2+}$, i.e. $\alpha$-particles. If it is electrons or He ions that are counted, the resulting $\alpha$-particles do not influence the observation. But in transmission, those additional photon absorptions do contribute to the counting. The calculation must therefore include that additional channel of photon absorption, for which the cross section is ${1.2}\times{10^{-18}}$cm${}^2$; about the same as the one for the single-photon ionization of the neutral, at the smooth part of the continuum, away from the resonance.
For the sake of completeness, we have included that additional channel in our calculations, by writing 
\begin{equation}
\dot P_{ion}^\diamondsuit (t) = {\dot P_{ion}}(t) - P_{ion}^\diamondsuit (t){\gamma _{DI}}
\end{equation}
where ${P_{ion}^\diamondsuit (t)}$ is the ionization probability including the double ionization of Helium and ${{\gamma _{DI}}}$ is the rate with which the Helium ions  produced from autoionization, absorb one additional photon. This rate can be expressed as the product of the relevant cross section of the process and the photon flux. The results will be compared with the numerical results of Appendix A, in which we use equations (18) and (19) under the assumption of a Gaussian shaped pulse.

For the Helium 2s2p ${}^1P$ AIS, the parameters involved in the theory (expressed in atomic units) are \cite{ref20,ref21,ref22}: $q = -2.79$, $\Gamma  = 1.37 \times {{10}^{ - 3}}$, $\Omega  = 0.025\frac{{{E_0}}}{2}$, $\gamma  = 0.1775{{\rm I}_0}$. We also set ${\omega _g} = 0$, therefore the energy difference ${\omega _{ag}}$ ($\hbar  = 1$) is equal to the energy of the 2s2p ${}^1P$  AIS, namely ${\omega _{ag}} = 65.40eV \simeq 2.211$a.u. The autoionization lifetime is approximately 18fs.

\begin{figure}[!ht]
	\centering
		\includegraphics[width=9cm]{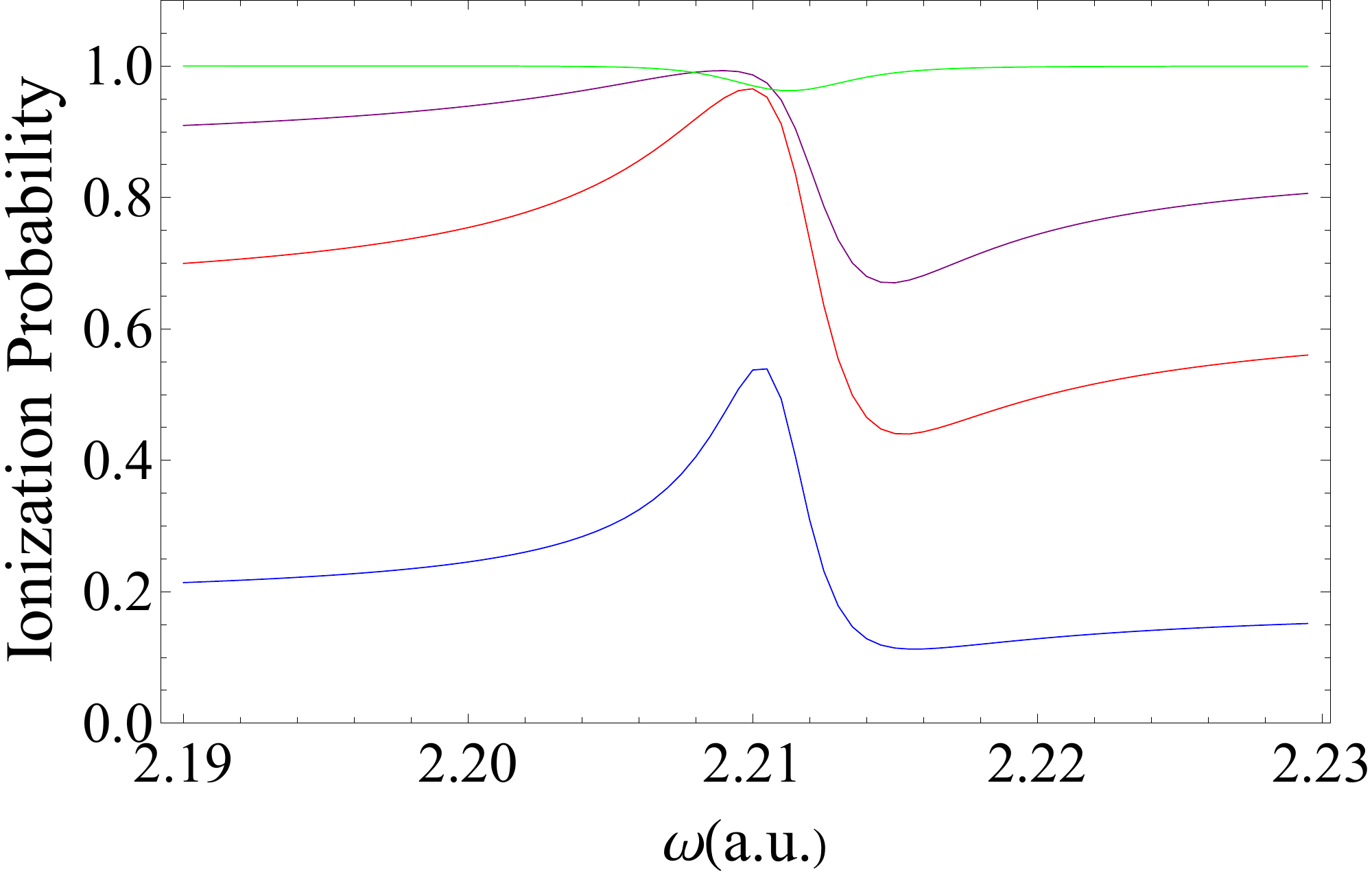}
	\caption[figure1]{Probability of ionization as a function of the driving frequency for various intensities and $T = 120fs$ , ${\gamma _L} = 0.0018$a.u. Blue Line: ${I_0} = {10^{13}}$W/cm${}^2$, Red Line: ${I_0} = 5 \times {10^{13}}$W/cm${}^2$, Purple Line: ${I_0} = {10^{14}}$W/cm${}^2$ and Green Line: ${I_0} =5 \times {10^{14}}$W/cm${}^2$.}
\end{figure}

Figure 1 illustrates the effects of the intensity on the AI profile. Notice that the Rabi frequency becomes comparable to the AI width at intensities around $2 \times {10^{14}}$W/cm${}^2$. For an interaction time around 120fs we see that, for strong intensities the ionization profile is almost flat. For weaker intensities the profile has an asymmetric form with its peak around the resonant frequency. The minimum, that is also barely visible in the Green curve, arises due to the interference between the direct ionization channel and the indirect channel via the configuration interaction, i.e. the Coulomb interaction for the problem at hand. The position of the minimum is intensity-dependant, a result that has also been noted before \cite{ref5}, due to the modification of the interference as the intensity  changes. In contrast to the non-fluctuating weak field case \cite{ref14}, we can see that the ionization probability is non-zero at the minimum. This result is due to the presence of a finite laser bandwidth, which samples signal from the wings although the driving frequency is tuned exactly at the minimum. As a result, the minimum is "filled out". Even for weak intensities the profile bears no resemblance to the usual textbook Fano profile \cite{ref14} and an attempt to fit the weak field curves with the standard Fano parameters $q$ and $\varepsilon$, leads to totally irrelevant values of $q$.

\begin{figure}[!ht]
	\centering
		\includegraphics[width=9cm]{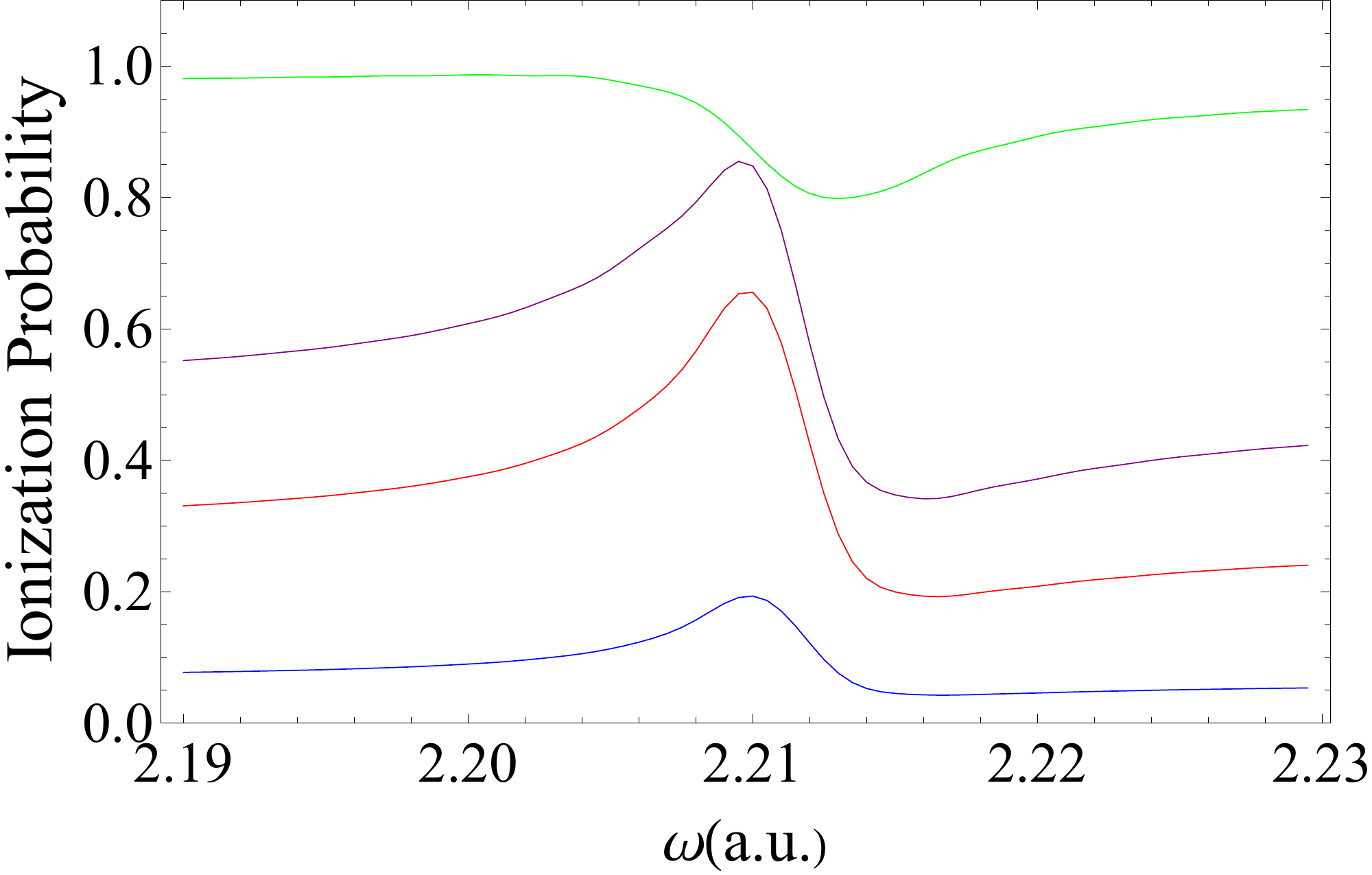}
	\caption[figure2]{Probability of ionization as a function of the driving frequency for various intensities and $T = 20$fs, ${\gamma _L} = 0.0018 $a.u. Blue Line: ${I_0} = {10^{13}}$W/cm${}^2$, Red Line: ${I_0} = 5 \times {10^{13}}$W/cm${}^2$, Purple Line: ${I_0} = {10^{14}}$W/cm${}^2$ and Green Line: ${I_0} =5 \times {10^{14}}$W/cm${}^2$.}
\end{figure}
It might seem reasonable to infer that the distortion of the profile in high intensities is due to power broadening. Although the intensity does play a significant role in the modifications of the AI profile, it is the combination of intensity and interaction time that truly determines the profile shape. In figure 2, we choose an interaction time that is 6 times smaller, i.e. $T = 20$fs for different values of the intensity. At intensities such that the Rabi frequency is comparable to the AI width, we can see that the profile is not completely flat if the interaction time is chosen so that it is comparable to the AI lifetime.

\begin{figure}[!ht]
	\centering
		\includegraphics[width=9cm]{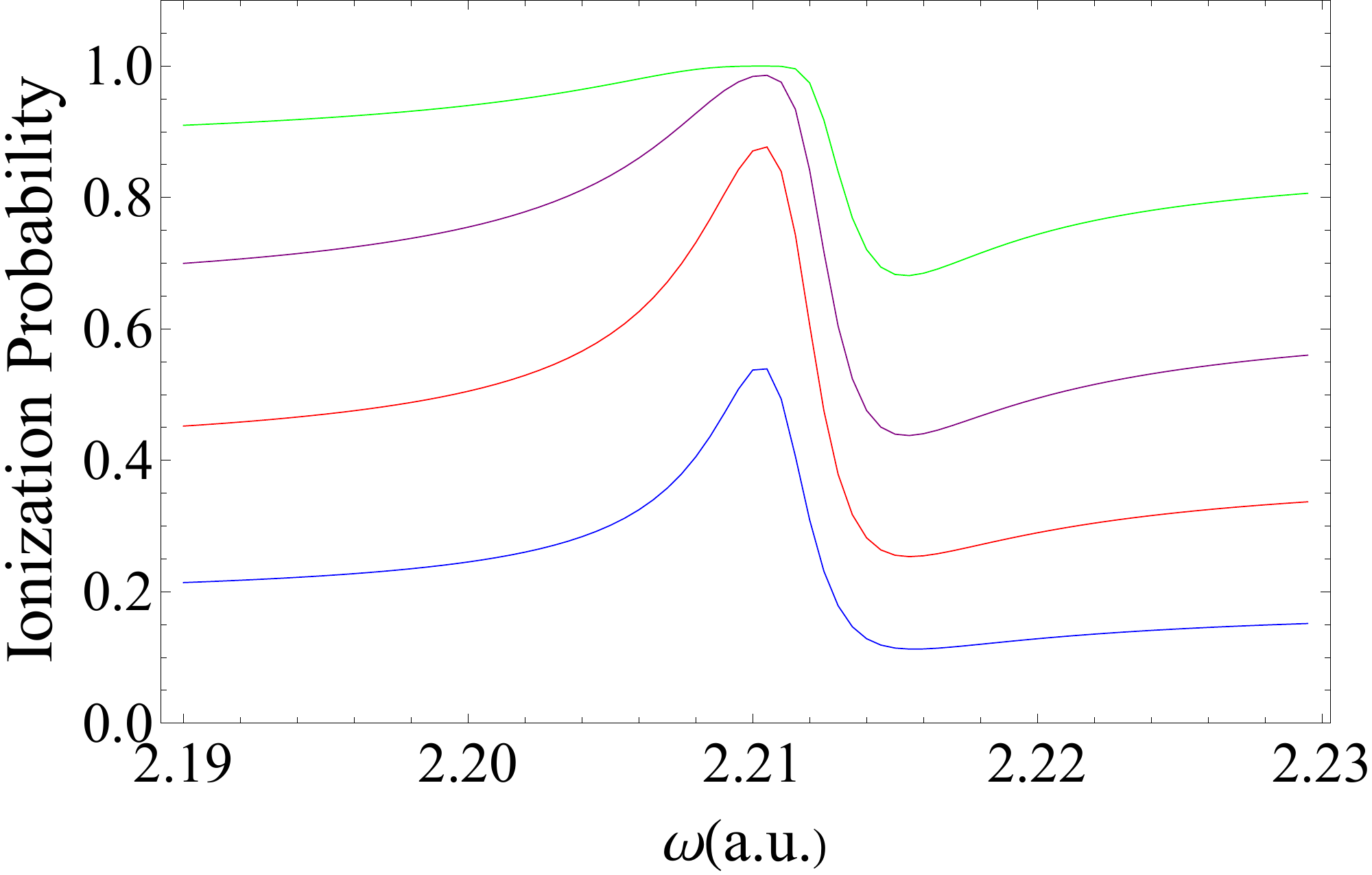}
	\caption[figure3]{Probability of ionization as a function of the driving frequency for various interaction times T and ${I_0} = {10^{13}}$W/cm${}^2$, ${\gamma _L} = 0.0018$a.u. Blue Line: $T = 120$fs, Red Line: $T = 240$fs, Purple Line: $T = 480$fs and Green Line: $T = 960$fs.}
\end{figure}

In figure 3 we explore the effects of the interaction time on the AI profile for a weak field of constant intensity. The interactions times are chosen sufficiently larger compared to the AIS lifetime (18fs). As the interaction time increases, the ionization probability generally increases, as there is more time available for the atom to be ionized. From figures 1 to 3 we can safely assume that the ionization is mainly determined by whether the system is time saturated, i.e. if for a given intensity, the time that the field is present is sufficient for the atom to be ionized completely.  

\begin{figure}[H]
	\centering
		\includegraphics[width=9cm]{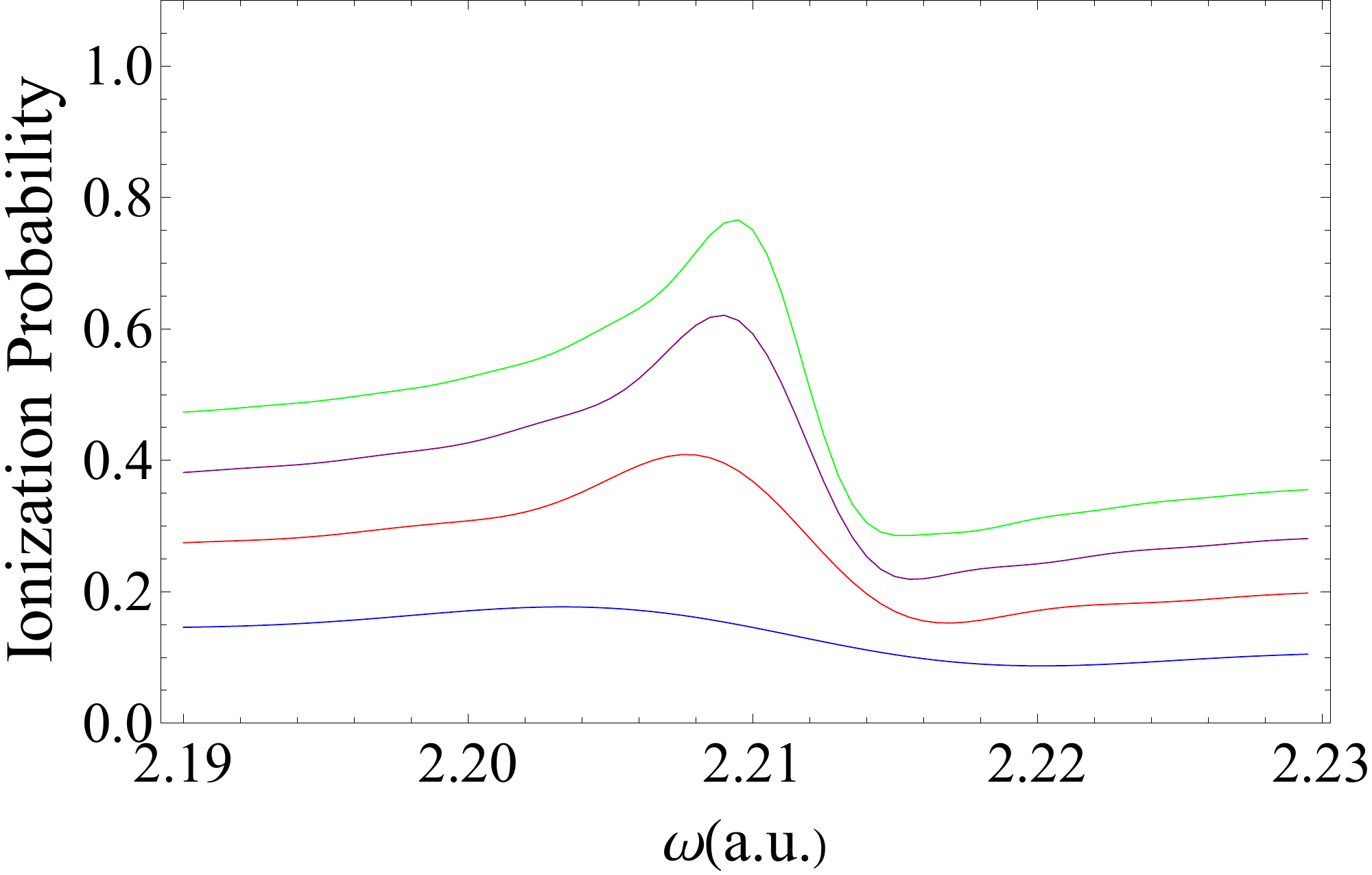}
	\caption[figure4]{Probability of ionization as a function of the driving frequency for various interaction times T and ${I_0} = {10^{14}}$W/cm${}^2$, ${\gamma _L} = 0.0018$a.u. Blue Line: $T = 5$fs, Red Line: $T = 10$fs, Purple Line: $T = 15$fs and Green Line: $T = 20$fs.}
\end{figure}

If on the other hand, the interaction time becomes sufficiently short, we will begin to observe a broadening of the profile, with the asymmetry obscured. As an example,  in figure 4 the interaction times are chosen such that they introduce a visible Fourier broadening. The main reason of this distortion is not the Fourier broadening itself, but a combination of power broadening and Fourier broadening due to short pulse durations, with the last being the dominant broadening mechanism.

\begin{figure}[H]
	\centering
		\includegraphics[width=9cm]{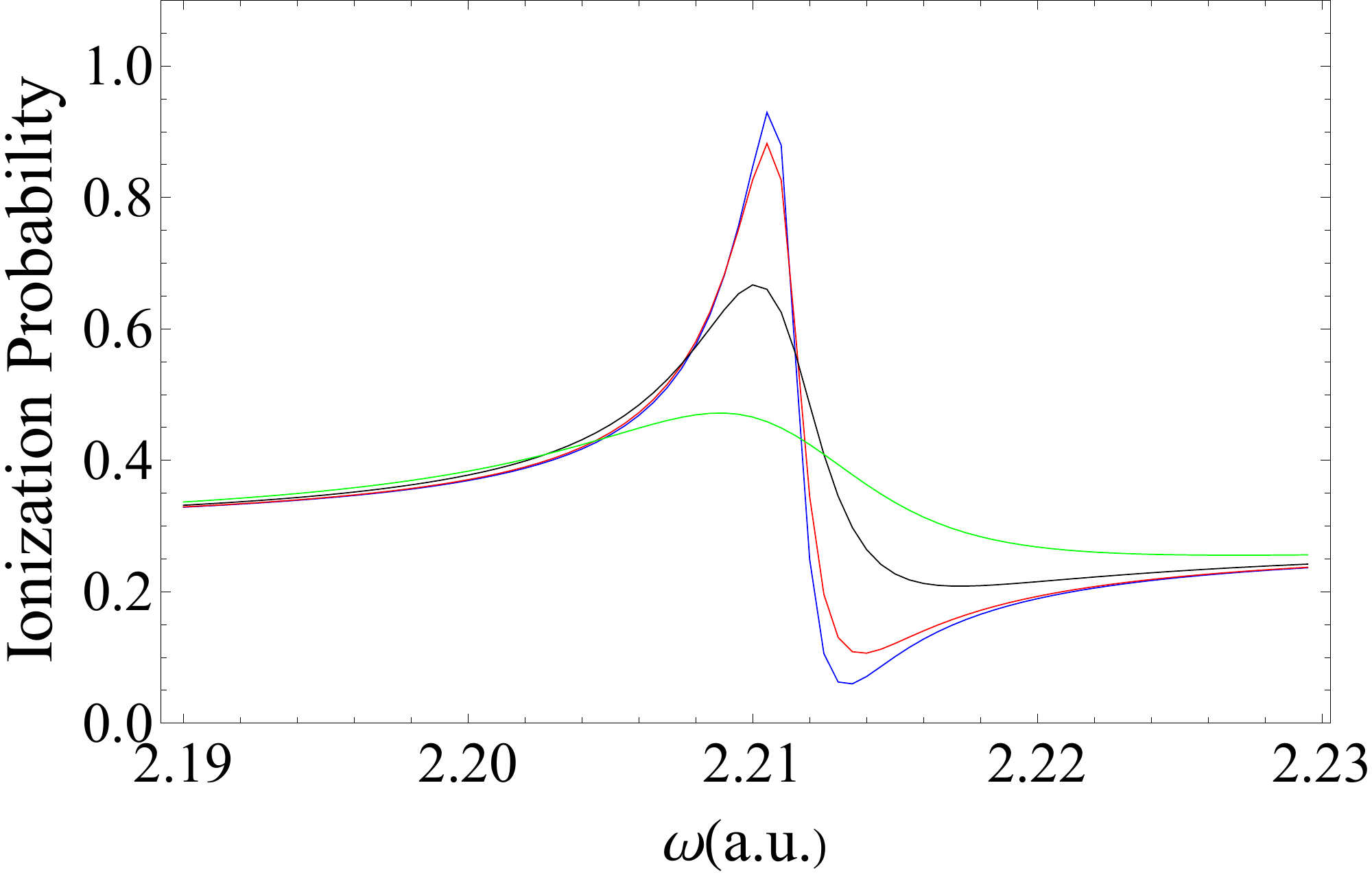}
	\caption[figure4]{Probability of ionization as a function of the driving frequency for various laser bandwidths and ${I_0} = {10^{13}}$W/cm${}^2$, $T = 150$fs. Blue Line: ${\gamma _L} = 0.0001$a.u. , Red Line: ${\gamma _L} = 0.0005 $a.u., Black Line: ${\gamma _L} = 0.003$a.u. and Green Line: ${\gamma _L} = 0.001$a.u.}
\end{figure}

In figure 5 we illustrate the effects of the field bandwidth on the AI profile. We can clearly see that when the bandwidth of the field becomes sufficiently larger than the autoionization width $\Gamma  = 1.37 \times {{10}^{ - 3}}$ a.u., the FWHM of the profile is mainly determined by ${\gamma _L}$. As the bandwidth increases the profile broadens and both the minimum and the maximum tend to be smoothed out. For very large bandwidths the asymmetry tends to become less visible since the profile becomes almost flat. If the bandwidth is sufficiently smaller than the AI width then the FWHM is determined mainly by $\Gamma$ or by the interaction time if it is shorter than the AI lifetime.

\section{\label{sec:leve14}Closing Remarks}

The effects discussed in the previous sections were of no importance in experiments under synchrotron sources which have small bandwidths and low intensity. However, the theory shows that when the laser bandwidth is substantial and is included in the calculations, the resulting profiles can be distorted dramatically. At the same time parameters such as pulse duration and intensity are interwoven in a non-linear fashion, leading to further distortion of the profile. For intensities in the strong field regime, as defined above, the presence of intensity fluctuations ushers in problems that we have only glimpsed at in this paper. The theoretical techniques developed and employed in previous work \cite{ref2} in the context of bound states are not directly applicable here. One way of approaching that case is through numerical simulation, as employed in \cite{ref19} for Auger resonances. An AI resonance, however, presents additional difficulties due to the interference between the discreet-discreet and discreet-continuum transitions. Although one might reasonably expect the qualitative behavior to be similar to that found for Auger resonances, substantial quantitative differences are to be expected. The strong driving of AI resonances viewed over the last 30 years is never free of surprises. The strong coupling between AI resonances \cite{ref4,ref5,ref6} in the presence of intensity fluctuations promises to be even more challenging. In view of ongoing and planned related experiments under FEL radiation, these theoretical issues have now come to center stage.

\section*{Acknowledgements}
The authors would like to thank G. M. Nikolopoulos for his careful reading of the manuscript and helpful remarks.
We also gratefully acknowledge many and ongoing discussions with Drs. Thomas Pfeifer and Christian Ott, on experimental as well as theoretical issues related to this work. 

\clearpage

\section*{Appendix: Gaussian Pulse Shape Results}
In this appendix we present our results using equations (18) and (19) for a Gaussian shaped pulse. For the sake of comparison with the analytical results of constant intensity, we use the same combinations of the relevant parameters appearing in figures 1 to 5. Note that for a Gaussian pulse we refer to ${I_0}$ as the peak intensity.

\begin{figure}[H]
	\centering
		\includegraphics[width=9cm]{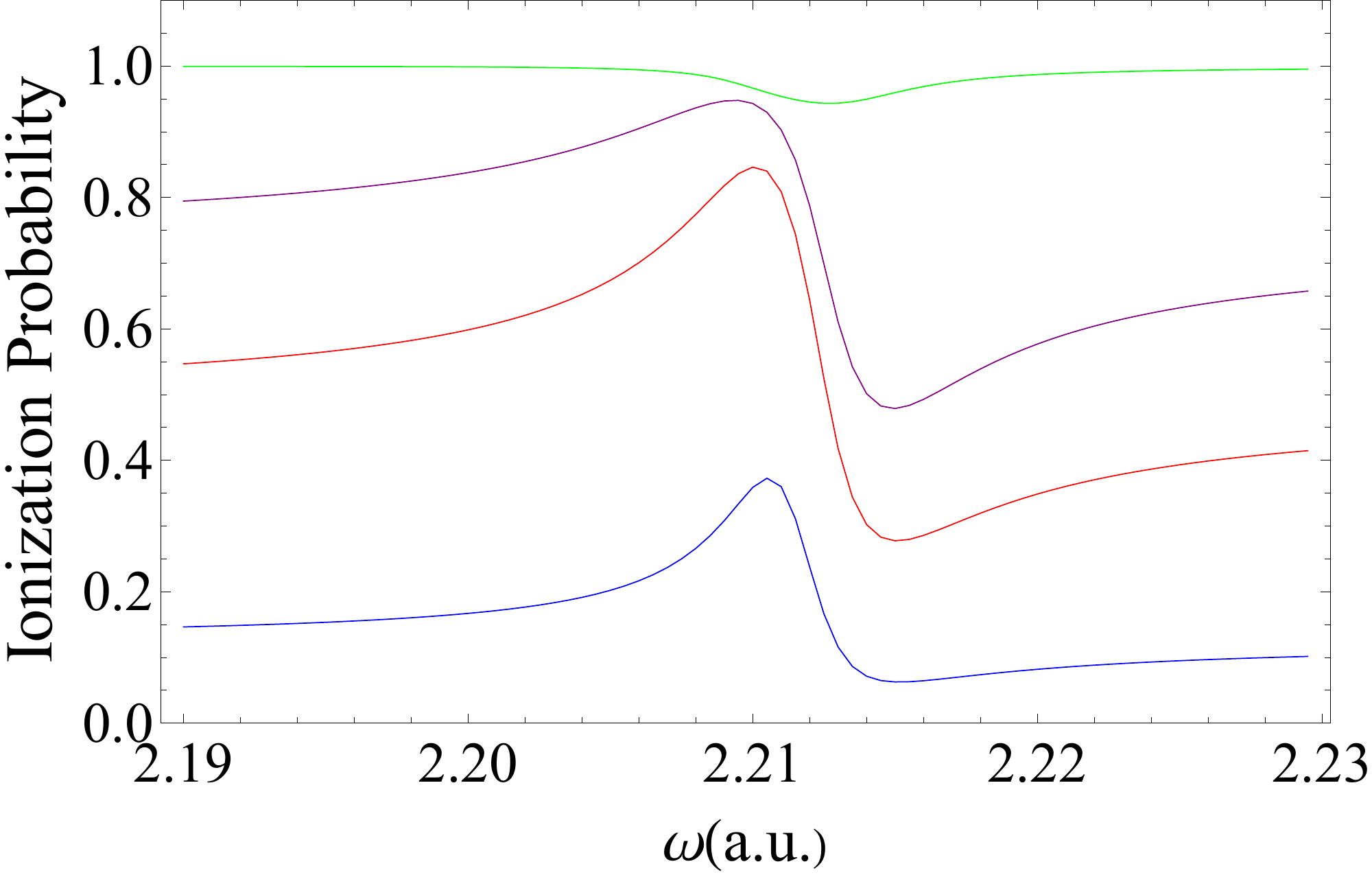}

	\caption[figureA6]{Probability of ionization as a function of the driving frequency for various intensities and $T = 120fs$ , ${\gamma _L} = 0.0018$a.u. under a Gaussian pulse. Blue Line: ${I_0} = {10^{13}}$W/cm${}^2$, Red Line: ${I_0} = 5 \times {10^{13}}$W/cm${}^2$, Purple Line: ${I_0} = {10^{14}}$W/cm${}^2$ and Green Line: ${I_0} =5 \times {10^{14}}$W/cm${}^2$.}
\end{figure}

\begin{figure}[H]
	\centering
		\includegraphics[width=9cm]{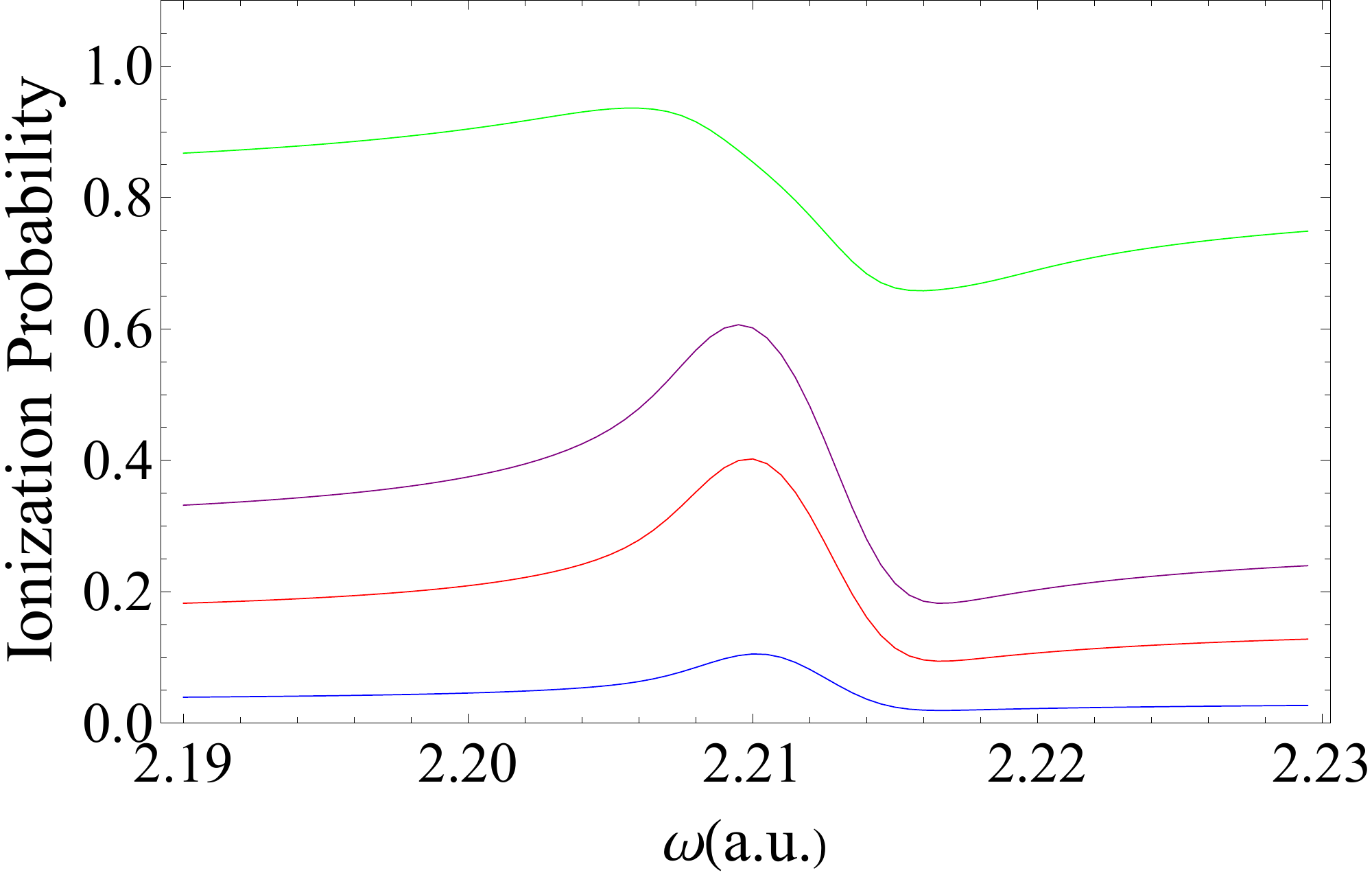}
	\caption[figureA7]{Probability of ionization as a function of the driving frequency for various intensities and $T = 20$fs, ${\gamma _L} = 0.0018 $a.u. under a Gaussian pulse. Blue Line: ${I_0} = {10^{13}}$W/cm${}^2$, Red Line: ${I_0} = 5 \times {10^{13}}$W/cm${}^2$, Purple Line: ${I_0} = {10^{14}}$W/cm${}^2$ and Green Line: ${I_0} =5 \times {10^{14}}$W/cm${}^2$.}
\end{figure}

\begin{figure}[H]
	\centering
		\includegraphics[width=9cm]{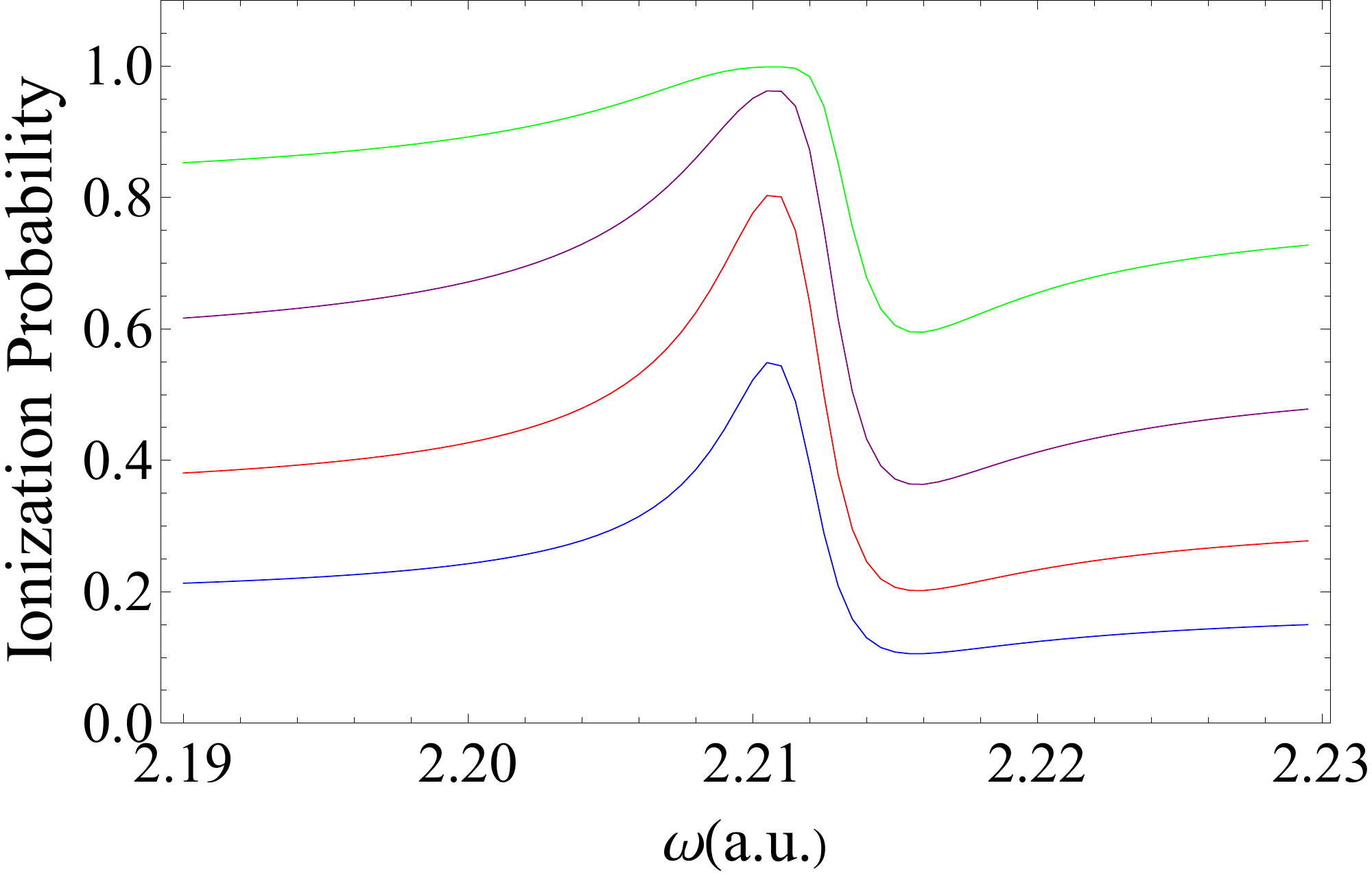}
	\caption[figureA8]{Probability of ionization as a function of the driving frequency for various interaction times T and ${I_0} = {10^{13}}$W/cm${}^2$, ${\gamma _L} = 0.0018$a.u. under a Gaussian pulse. Blue Line: $T = 120$fs, Red Line: $T = 240$fs, Purple Line: $T = 480$fs and Green Line: $T = 960$fs.}
\end{figure}

\begin{figure}[H]
	\centering
		\includegraphics[width=9cm]{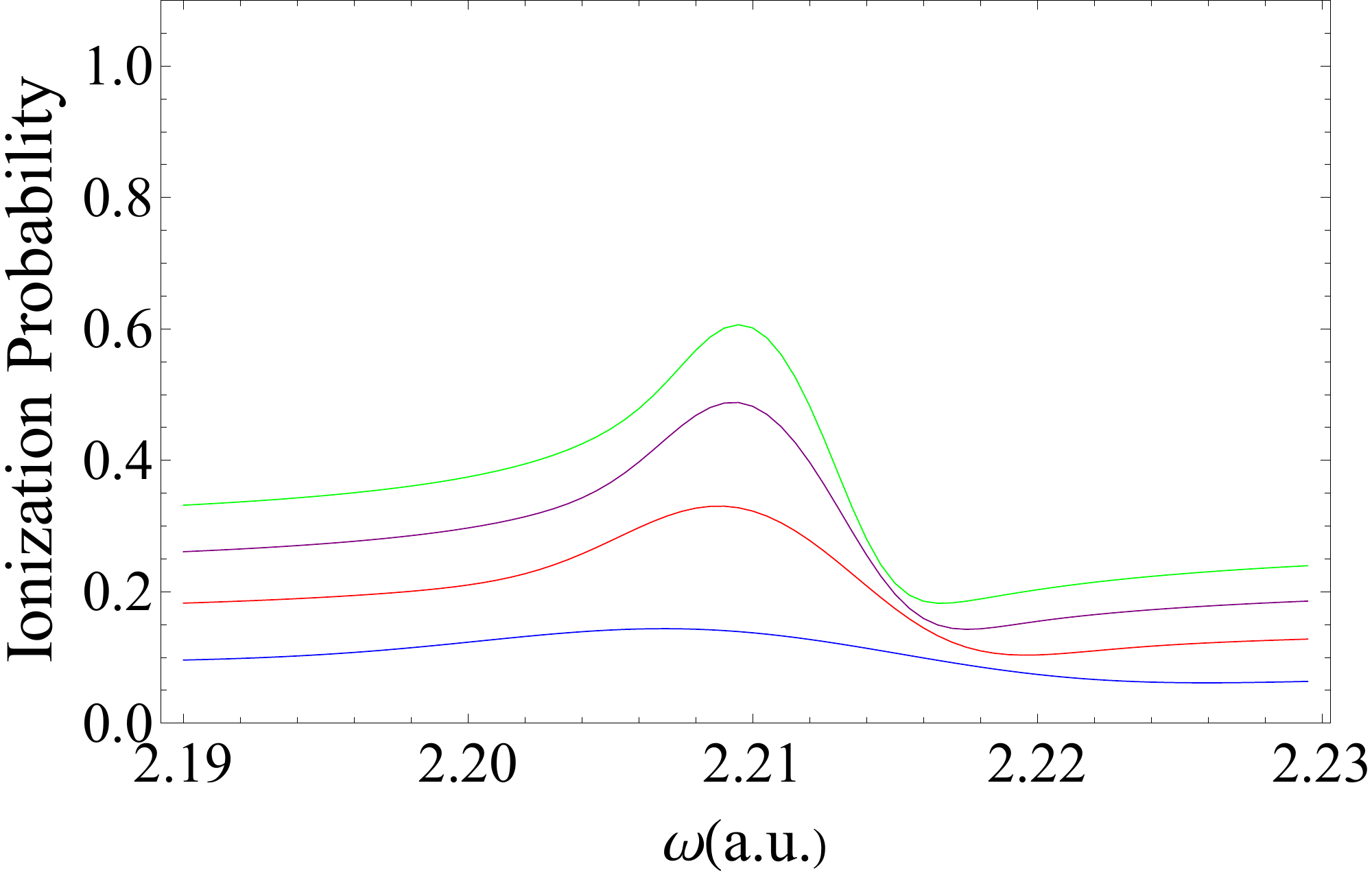}
	\caption[figureA9]{Probability of ionization as a function of the driving frequency for various interaction times T and ${I_0} = {10^{14}}$W/cm${}^2$, ${\gamma _L} = 0.0018$a.u. under a Gaussian pulse. Blue Line: $T = 5$fs, Red Line: $T = 10$fs, Purple Line: $T = 15$fs and Green Line: $T = 20$fs.}
\end{figure}

Comparing the above diagrams to the ones presented in the previous section we can safely accept that the shape of the pulse generally does not have a very important impact on the determination of the AI profile. This result is generally known and is widely used for analytical and numerical simplifications. However, the comparison between figures 2 and 7 as well as 4 and 9 respectively, reveals that if the pulse duration is of the order of the AI lifetime, a Gaussian pulse produces  a lineshape different from that produced by a square pulse, under the same combination of the relevant parameters. The general picture that arises is that the Gaussian pulse for such interaction times is less effective in AIS ionization than a square pulse. This seems quite logical since the square pulse forces the atom to be driven by a field whose intensity is ${I_0}$ over the whole duration of the pulse, whereas the Gaussian pulse has a peak value of ${I_0}$ and its wings at smaller intensities. The above results have also been tested with trapezoidal and Lorentzian pulse shapes and the picture doesn't differ a lot. However, we should always be aware that equations (18) and (19) are rate equations and the validity of the results becomes questionable  for very strong fields where the approximation (17) breaks down.

\begin{figure}[H]
	\centering
		\includegraphics[width=9cm]{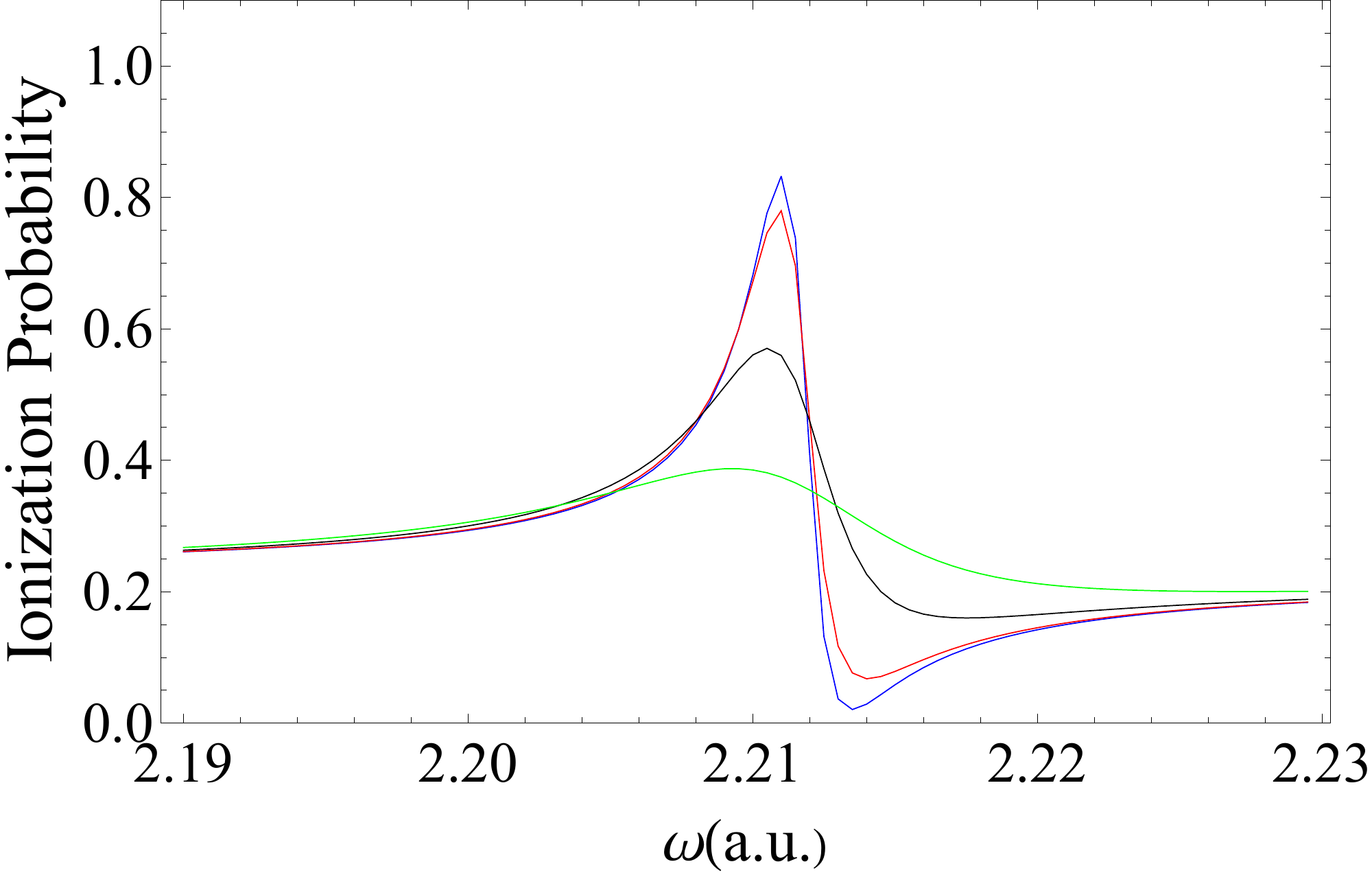}
	\caption[figureA10]{Probability of ionization as a function of the driving frequency for various laser bandwidths and ${I_0} = {10^{13}}$W/cm${}^2$, $T = 150$fs under a Gaussian pulse. Blue Line: ${\gamma _L} = 0.0001$a.u. , Red Line: ${\gamma _L} = 0.0005 $a.u., Black Line: ${\gamma _L} = 0.003$a.u. and Green Line: ${\gamma _L} = 0.001$a.u.}
\end{figure}

\end{document}